\documentstyle[12pt]{article}
\newcommand{\be}{\begin{equation}}
\newcommand{\ee}{\end{equation}}

\newcommand{\la}[1]{\label{#1}}
\newcommand{\r}[1]{(\ref{#1})}
\begin{document}
\title{Isgur-Wise functions for confined light quarks \\ in a colour electric
potential\thanks{Work supported in part (M.S.) by the KBN
grant no.\ 20-38-09101. We also thank the Norwegian Research Council for a
travel grant.}
}
\author{
{\sc H.\,Ho\hspace{-0.3cm}/gaasen}\thanks{E-mail {\sc hogaasen@vuoep6.uio.no}}
 \\
{\it Department of Physics, University of Oslo,} \\
{\it N-0316 Oslo, Norway.}
\vspace{0.5cm}\\
and
\vspace{0.5cm}\\
{\sc M.\,Sadzikowski}\thanks{E-mail {\sc ufsadzik@ztc386a.if.uj.edu.pl}}\\
{\it Institute of Physics, Jagellonian University,} \\
{\it ul.\,Reymonta 4, PL-30\,059 Krak\'ow, Poland.}}
\date{}

\maketitle
\begin{center}

\begin{abstract}

We explore the influence on the Isgur-Wise function of the colour electric
potential between heavy and light quarks in mesons.
It is shown that in bag models, its inclusion tends to restore light quark
flavour symmetry relative to the MIT bag predictions, and that relative to this
model it  flattens the Isgur-Wise function.
Results compare very well with observations.

\end{abstract}
\end{center}
\newpage


The physics of states containing a heavy quark is the subject of much research.
This is evident because in the limit where the mass of the heavy quark grows
beyond any limit, the heavy quark can be regarded as a static source where the
heavy quark spin effectively decouples from the interaction. The degrees of
freedom of the light quark(s) then become the main dynamical variables, and the
system is a fine theoretical laboratory to study properties of bound light
quarks.
It is hoped that the properties one can deduce for such systems is of relevance
for physical systems where the heavy quark is a $b$ or a $c$ quark.

One of the most brilliant applications of the Heavy Quark Symmetries
(see for example \cite{Neub} and citation therein) is the
description of semileptonic decays of particles containing a heavy quark
in terms of only one unknown quantity the Isgur - Wise function (IW)
\cite{IW}.
The dependence of this function of light degrees of freedom is very
complicated, and establishes an essential problem in all calculations
that leads to numerical values of, not only branching ratios of
semileptonic decays, but also in part of two-body decays branching ratios
and determinations of the element $|V_{cb}|$ of Kobayashi-Maskawa matrix.

A reliable calculation of the IW function based on  first
principles is beyond our power up to now, so a variety of different
models must be used to estimate some numerical results.
One of the possible approaches in the calculation of the IW function
was based on
the MIT bag model \cite{sadzal}. Bag models were successful in describing
several important features connected with particles containing one heavy
quark (for example in papers \cite{Shuryak}, \cite{Izatt}, \cite{Wilcox},
\cite{Sad}). In this paper we deal with the Coulomb-like bag model, a very
natural candidate for reasonable description of heavy - light quark system.

In the infinite mass limit the Heavy Quark Effective Theory (\cite{Neub})
predicts that the IW function is related to the overlap of the
light degree of freedom wavefunction multiplied  by a kinematical factor.
In the case of the ground state to ground state transitions we have
\cite{Zal}\footnote{In case of $\Lambda $ baryons the IW function is often
defined in literature also as a pure overlap, without kinematical factor.}:
\be
\xi (\omega ) = \sqrt{\frac{2}{\omega + 1}} <\Phi_{Q^{'}l}|\Phi_{Ql}>
\la{ksi}
\ee
where the ket in above equation describes the light degree of freedom
wavefunction of the parent ($\Phi_{Ql}$) and the bra of the
daughter ($\Phi_{Q^{'}l}$) particle. The IW function is independent of the
spin of the heavy quark.
Before we come to our main topic we make some general remarks
about the IW function valid for all models satisfying the following  two
 assumptions:
\begin{itemize}
\item The valence quark approximation
\item Spherical symmetry of the ground state probability density
\end{itemize}
Due to the first assumption the wavefunction of the light quark in the
particle rest frame can always be written in the form:
\be
\Phi^{0}(x) = \Phi^{0}(\vec{x})e^{-iEt}
\la{fi}
\ee
where $\Phi^{0}(\vec{x})$ is a bispinor field and $E$ is its ground state
energy.\\
The second assumption states that the product:
\be
\rho (r) \equiv \Phi^{0 \dag }(x)\Phi^{0}(x)
\la{ro}
\ee
is independent of angles.

Consider the IW function for mesons. In the decay of one particle to another
one can always choose the frame where two particles move along the z-axis
with equal and opposite velocities (modified Breit frame \cite{Hog}). In this
frame the overlap takes the form:
\be
<\Phi_{Q^{'}l}|\Phi_{Ql}> = \int_{CK} d^{3}x \Phi^{\dag }_{Q^{'}l}(x)
\Phi_{Ql}(x) |_{t=0}
\la{ov}
\ee
where:
\be
\Phi_{Ql}(x) = S(-\vec{v})\Phi^{0}_{Ql}(x,y,\gamma z)e^{-iE\gamma vz}
\la{fibust-}
\ee
is a wavefunction describing the light quark in the parent particle that is
moving with velocity $v=|\vec{v}|$ along the negative z-axis.
\be
\Phi_{Q^{'}l}(x) = S(\vec{v})\Phi^{0}_{Q^{'}l}(x,y,\gamma z)e^{iE\gamma vz}
\la{fibust+}
\ee
is a wavefunction describing the light quark of the daughter particle which is
moving with velocity $v$ in positive direction along the z-axis.\\
Both functions are related to the wavefunction in their rest frames via
boost operator $S(\vec{v}$) and Lorentz transformations:
\be
L^{-1}_{\pm \vec{v}}(0,\vec{x})=(\mp\gamma vz,x,y,\gamma z)
\la{loretz}
\ee
We define the volume $CK$ as the region explored by both wavefunctions in the
moment of the decay. In case of the MIT bag model it is the intersection of
contracted bags \cite{sadzal}, in case of other models it may be the
"infinite" region.  Lets also note, that according to Heavy Quark Symmetries
the wavefunctions $\Phi^{0}_{Q^{'}l}$ and $\Phi^{0}_{Ql}$ are
the same.

If one writes down the overlap in terms of the wavefunctions from their
rest frames, then in the integral over space there will always appear a
product  of boost matrices:
$S^{\dag }(\vec{v})S(-\vec{v})$. Using the identity between the
velocity $|\vec{v}|$ and scalar product of velocities of both particles
$\omega =vv^{'}$ valid in the "Breit frame":
\be
|\vec{v}|=\sqrt{\frac{\omega - 1}{\omega + 1}}
\la{ident}
\ee
one can write:
\be
S^{\dag }(\vec{v})S(-\vec{v})=\sum^{\infty }_{n=0} \hat{\alpha }_{n}
(\frac{\omega - 1}{\omega + 1})^{n}
\la{rozkl}
\ee
where only the first operator $\hat{\alpha}_{0}$ is known. It is equal to
unity. Using \r{ksi}, \r{ov}, \r{fibust-}, \r{fibust+} and \r{rozkl} and
rescaling the z - variable by a factor $\gamma $ one gets:
$$
\xi (\omega ) = (\frac{2}{\omega +1})[\int_{K(0,R)} d^{3}x \rho (r)
j_{0}(2E\sqrt{\frac{\omega - 1}{\omega + 1}}r) \, +
$$
\be
\la{ks1}
\ee
$$
+ \sum_{n=1}^{\infty }(\frac{\omega - 1}{\omega + 1})^{n}\int_{K(0,R)}d^{3}x
\Phi^{0 \dag }(\vec{x})\hat{\alpha }_{n}\Phi^{0}(\vec{x})
e^{-2iEvz}]
$$
where $K(0,R)$ is the region explored by the light degree of freedom
wavefunction in the rest frame of particle. For the simplest bag models
it is a spherical bag but for other models it can be an "infinite" region.
$j_{0}$ is the usual spherical Bessel function of order zero. Now it is
straightforward to find the slope parameter:
\be
\rho^{2} = -\frac{d\xi }{d\omega }|_{\omega =1} = \frac{1}{2} +
\frac{1}{3}E^{2}<r^{2}> - \frac{1}{2}\int_{K(0,R)} d^{3}x
\Phi^{0 \dag }(\vec{x})\hat{\alpha }_{1}\Phi^{0}(\vec{x})
\la{ro1}
\ee
Unfortunately these results are not giving us enough information.
The additional assumption that $S(\vec{v})$ is hermitian
(for example true for the free boost) leads to the following formula:
\be
\hat{\alpha }_{n}(r) = 0 \,\,\,\,\,\,\,\,\,\,\, for \,\,\,\,\, n=1,2,...
\la{zal2}
\ee
and consequently:
\be
\xi (\omega ) = (\frac{2}{\omega +1})\int_{K(0,R)} d^{3}x \rho (r)
j_{0}(2E\sqrt{\frac{\omega - 1}{\omega + 1}}r)
\la{ksi3}
\ee
\be
\rho^{2} = \frac{1}{2} + \frac{1}{3}E^{2}<r^{2}>
\la{ro3}
\ee
These relations were found in paper \cite{sadzal}.
The above results have a practical
meaning for all models in which the density and the ground state energy of the
light degrees of freedom are known.


It was shown earlier by K. Zalewski and one of us \cite{sadzal}, that in the
case of mesons the MIT bag model wave functions for light quarks gave results
for the IW function that compared very well with what we believe is known
experimentally.
When we speak of the MIT bag model here we speak of its simplest tractable
version, the static approximation where the bag is spherical in its rest frame.
The confinement mechanism then correspond to a world scalar potential which is
zero inside and infinite outside the bag surface. There is no four vector like
potential between the "infinitely" heavy quark, which defines the center of
the bag, and the light quark. Up to maximal distance the bag radius, the light
quark satisfies the free particle Dirac equation.
One certainly would believe that there is an influence of the gluonic
interaction at intermediate distances, theoretically one would expect
something
like a Coulomb-like potential, modified at very small distances by asymptotic
freedom effects. This certainly is the result from lattice gauge calculations
of the potential between two heavy quarks.
It is therefore of some interest to find out how such an inner potential
influences the successful prediction of the IW function that was obtained
 from the  simplest (MIT) bag wave functions for the light quark.
We therefore model the interaction of the heavy and light quark with a Coulomb
potential for $r < R$, solve the Dirac equation there and quantize the quark
energy with the MIT-Bogoliubov boundary condition at $r=R$ \cite{hogm}.
We normalize the potential so that it is zero at the boundary. We shall
comment later on the influence of asymptotic freedom effects at very small
$r$. The inner potential is now of the form:
\be
   V(r)=-b \left(\frac{1}{r}-\frac{1}{R}\right)
\la{V}
\ee
We now define the light quark energy $E$ as solution of the Dirac equation:

\be
H\Phi^{0}(r)=E\Phi^0(r)
\la{phi}
\ee
where:
\be
H=\vec{\alpha }\vec{p} + \beta m + V(r)
\la{H}
\ee
Here $m$ is the mass of the light quark.
The quark energy is determined from the boundary condition at the bag radius
$R$:
\be
i\vec{\gamma }\hat{r}\Phi^{0}(R) = \Phi^{0}(R)
\la{bc}
\ee
where $\hat{r}$ is a radial unit vector.
The light quarks wave function and its energy are evidently functions of the
strength of the central Coulomb potential characterized by the constant $b$.
This is of course not a parameter that we know the size of. For illustrative
purposes we shall use the value obtained from a fit to charmed mesons in
\cite{Wilcox}, $b=0.542$ and their corresponding bag radii for strange and
nonstrange quarks. This value of $b$ correspond to a strong
coupling constant $\alpha_{s} = 0.407$, a quite reasonable value.
All parameters now fixed, we easily compute the IW function from formula
\r{ksi3}.

In fig. 1, we show the IW function for massless light quarks in the cases
$b=0$ (MIT bag \cite{sadzal}), $b=0.542$ \cite{Wilcox} and the ARGUS data
\cite{argus}.
To compare the IW function with the data we fit the Cabbibo-Kobayashi-Maskawa
matrix parameter $|V_{cb}| = 0.0386 \sqrt{\tau_{B}/1.29ps}$. The $\chi^2$ of
this fit
is 9.3 for 7 degrees of freedom. As the IW functions we used the approximation
of the exact result by the functions:
\be
\xi^{B}(\omega ) = (\frac{2}{\omega +1})^{1.52+\frac{0.45}{\omega }}
\la{ksib}
\ee
for $\bar{B} \rightarrow De\bar{\nu_{e}}$ and
\be
\xi^{B_{s}}(\omega ) = (\frac{2}{\omega +1})^{1.67+\frac{0.59}{\omega}}
\la{ksibs}
\ee
for $\bar{B_{s}} \rightarrow D_{s}e\bar{\nu_{e}}$ similar for ones used in
\cite{sadzal}.
The appropriate results for slope parameters are:
\be
\rho_{B}^2=0.98\,\,\,\,\,\,\,\,\,\,\,\, \rho_{B_{s}}^2=1.135
\la{rob}
\ee

The slope parameter for $\bar{B} \rightarrow De\bar{\nu_{e}}$ is in a good
agreement with what one finds in lattice calculations.
One group \cite{lat} found $\rho_{B}^2=1.0(8)$, another \cite{latt}
$\rho_{B}^2=1.2( +7 -3)$.

As can be seen from the figure, the differences between the free MIT bag
 and Coulomb-like bag
are not very big, but the central potential clearly has led to a flatter
IW function.

Considerable insight can be gained by looking at the expression for the
derivative of the IW function at minimum recoil, i.e. at $\omega=1$  given in
formula \r{ro3}, where the product X of the  quark energy and the root--mean
square radius of the light quark probability distribution enters in an
essential manner:
\be
X=E\sqrt{<r^{2}>}
\ee
In the region of interest to us this product is given to a very good
approximation by a linear fit
$X=a_1 + a_2mR$  where the coefficients $a_1$ and $a_2$ decreases with
increasing strength of the Coulomb potential $b$.
For $b=0$ we find:
\be
X=1.49+0.29mR
\la{er1}
\ee
for $b=0.542$:
\be
X=1.15+0.17mR
\la{er2}
\ee
This has interesting consequences that will show up in a moment, when we
compute decay probabilities: The Coloumb potential tend to restore flavour
symmetry in decay probabilities.

Using the formulas for decay rates (for example from \cite{sadzal}) and
the functions \r{ksib}, \r{ksibs} we get:
\be
Br(B \rightarrow Dl\bar{\nu }) = 1.79(\tau_{B}/1.29ps)\%
\la{bd}
\ee
\be
Br(B \rightarrow D^{\ast }l\bar{\nu }) = 5.05(\tau_{B}/1.29ps)\%
\la{bdstar}
\ee
\be
Br(B_{s} \rightarrow D_{s}l\bar{\nu }) = 1.74(\tau_{B_{s}}/1.29ps)\%
\la{bsds}
\ee
\be
Br(B_{s} \rightarrow D^{\ast }_{s}l\bar{\nu }) = 5.02(\tau_{B_{s}}/1.29ps)\%
\la{bsdsstar}
\ee
Agreement with available data ($1.7 \pm 0.4\% $) and ($4.8 \pm 0.6\% $)
\cite{pdg} for $B$ to $D$ and $B$ to $D^{\ast }$ transitions  is
quite satisfactory. What is more interesting we also found that:
\be
Br(B \rightarrow Dl\bar{\nu }) \approx Br(B_{s} \rightarrow D_{s}l\bar{\nu })
\la{eq1}
\ee
\be
Br(B \rightarrow D^{\ast }l\bar{\nu }) \approx
Br(B_{s} \rightarrow D^{\ast }_{s}l\bar{\nu })
\la{eq2}
\ee
as should be expected from naive observation based on $SU(3)$ flavour
symmetry. This result was absent in case of the MIT bag. It can be explained
by the fact that the increase of the energy of the ground state with the
increasing
light quark mass is much better compensated by the decrease of the mean
square radius in case of the Coulomb-like than in the MIT bag. This is the
source of the difference between \r{er1} and \r{er2}. The most important thing,
however, is that the product X is smaller in the Coulomb bag than in
the MIT bag so that the slope given by eq \r{ro3} does not vary as much with
varying light quark mass.

Measuring and comparing branching ratios to test light quark flavour symmetry
is therefore extremely interesting.

The sceptical reader might be worried by the fact that the Coulomb potential
does not incorporate asymptotic freedom, one of the most celebrated properties
of gluonic interaction. There are all reasons to check, whether the
singularity
of the Coulomb potential at the origin distorts our results. This is not the
case however: We integrated the Dirac equation with the potential $V(r)$ of
equation \r{V} for $r>r_{0}$ and  constant potential $V(r_{0})$ for $r<r_{0}$.
With $r_{0}=0.2$ $GeV^{-1}$ there were no differences on the $1\%$ level for
the calculated IW-function relative to the values we found with the pure
Coulomb potential.

\vspace*{0.5 cm}
\noindent
{\bf Acknowledgements}\\
We thank Kacper Zalewski for a useful discussion.

\end{document}